\documentclass[10pt,twocolumn,preprintnumbers,amsmath,amssymb,nofootinbib,superscriptaddress]{revtex4-2}

\usepackage[dvipsnames]{xcolor}
\usepackage{bm}
\usepackage{graphicx}
\usepackage[colorlinks,allcolors=RoyalBlue]{hyperref}

\newcommand{\eq}{\mathrm{eq}}
\newcommand{\mr}{\mathrm{mr}}
\newcommand{\noteq}{{\mkern 1.5mu\overline{\mkern-1.5mu \mathrm{eq}\mkern-1.5mu}\mkern 1.5mu}}

\newcommand{\kd}{\mathrm{kd}}
\newcommand{\dm}{\chi}
\newcommand{\sv}{\langle \sigma v \rangle}

\begin{document}

\title{Dark Matter Isocurvature from Curvature}

\author{Ian Holst, Wayne Hu, Leah Jenks}
\affiliation{Kavli Institute for Cosmological Physics, Department of Astronomy \& Astrophysics, Enrico Fermi Institute, The University of Chicago, Chicago, IL 60637, USA}

\begin{abstract}
Isocurvature fluctuations, where the relative number density of particle species spatially varies, can be generated from initially adiabatic, or curvature, fluctuations if the various species fall out of or were never in thermal equilibrium. The freezing of the thermal relic dark matter abundance is one such case, but for modes that are still outside the horizon the amplitude is highly suppressed and originates from the small change in the local expansion rate due to the local space curvature produced by the curvature fluctuation. We establish a simple separate-universe method for calculating this generation that applies to both freeze-in and freeze-out models, identify three critical epochs for this process, and give general scaling behaviors for the amplitude in each case: the freezing epoch, the kinetic decoupling epoch and matter-radiation equality. Freeze-out models are typically dominated by spatially modulated annihilation from the latter epochs and can generate much larger isocurvature fluctuations compared with typical freeze-in models, albeit still very small and observationally allowed by cosmic microwave background measurements. We illustrate these results with concrete models where the dark matter interactions are vector or scalar mediated.
\end{abstract}

\date{\today}

\maketitle

\section{Introduction} 

The primordial perturbations responsible for the large scale structure in our universe can be of two types: curvature (also known as adiabatic) or isocurvature. Generically, slow-roll, single-field inflation predicts only adiabatic perturbations, where the number density of particle species fluctuate together (e.g.~\cite{Wands:2000dp,Weinberg:2004kr}). More involved models such as multi-field inflation can generate isocurvature modes where the relative number density of species fluctuate as well (e.g.~\cite{Wands:2007bd}), though these modes are tightly constrained by cosmic microwave background (CMB) anisotropy \cite{Planck:2018jri}.

In principle, even if the primordial perturbations are adiabatic, the subsequent evolution of the universe can generate spatial fluctuations in the relative number density of species if thermal equilibrium between species is not maintained. One such possibility involves fluctuations between relic dark matter and the thermal radiation bath \cite{Bellomo:2022qbx}.

Dark matter is known to make up approximately a quarter of the energy density of the universe, however its specific properties and the manner of its production are yet unknown \cite{Rubin:1970zza,WMAP:2008lyn,Planck:2018vyg}. Two proposed mechanisms which can produce the correct relic abundance from the thermal radiation bath for a wide range of particle dark matter models are the freeze-in and freeze-out mechanisms. In the freeze-out mechanism the dark matter begins in thermal equilibrium with the Standard Model radiation bath. Once interactions that produce and destroy the dark matter become inefficient, its abundance is frozen, leaving behind a thermal relic that is no longer in equilibrium. In the freeze-in mechanism, dark matter interacts so weakly that it is never produced in equilibrium abundance; instead, it builds up slowly until the production channel is suppressed \cite{McDonald:2001vt, Bernal:2017kxu, Hall:2009bx}. Freeze-in and freeze-out represent the regimes at either end of a continuum of phenomena depending on the strength of interactions, but in both cases, the dark matter is out of equilibrium at the freezing epoch and beyond.

Recent investigations in the literature have raised the question of whether isocurvature fluctuations which arise from the dark matter freeze-in process are observable and ruled out by current CMB observations (\cite{Bellomo:2022qbx} v1). On the other hand, general perturbation theory and causal arguments imply that any such production can only produce isocurvature fluctuations that are suppressed on superhorizon scales by $k^2$, where $k$ is the comoving wavenumber \cite{Racco:2022svs, Strumia:2022qvj} (see also \cite{Bellomo:2022qbx} v2) but these arguments leave the exact amplitude and dynamics of the generation unspecified.

Our goal for this paper is to quantify the amplitude and clarify the mechanisms behind isocurvature generation from curvature fluctuations for the general class of thermal relic dark matter models. Because dark matter production is a local process, we can calculate its local abundance through the so-called separate universe formalism in which the effects of a long-wavelength perturbation on local small-scale observables can be absorbed into a change in the background cosmology, or ``separate universe" \cite{Sirko:2005uz,Gnedin:2011kj,Baldauf:2011bh, Li:2014sga, Hu:2016ssz,Zegeye:2021yml,Inomata:2023faq}. We use this separate universe approach to examine the connection between long-wavelength spatial curvature perturbations and dark matter isocurvature perturbations for both freeze-in and freeze-out. Our technique applies to any dark matter model where the final abundance depends on the local expansion rate and we illustrate the results for both freeze-in and freeze-out scalar and vector mediated models. 

Generically there are three epochs of interest which determine the final isocurvature amplitude: the freezing epoch, the kinetic decoupling epoch and the epoch around matter-radiation equality. Even though in all three cases the isocurvature generation is suppressed by $k^2$ outside of the horizon, the amplitudes can vary by many orders of magnitude for the same final relic dark matter abundance. In fact typical freeze-out models can have a much larger isocurvature amplitude than freeze-in models.

The structure of the paper is as follows: in \S\ref{sec:separate}, we review the separate universe formalism and develop the methodology for calculating isocurvature generation. In \S\ref{sec:models}, we describe our illustrative models for both freeze-in and freeze-out and in \S\ref{sec:results} present our numerical results for the isocurvature perturbations in each case. We conclude in \S\ref{sec:discussion} with a discussion of the general predictions for thermal relic dark matter and the observational implications of the very small isocurvature modes that result. Throughout this work we employ units where $\hbar=c=k_B=1$ and the scale factor $a=1$ at the present.

\section{Separate Universe Dark Matter Abundance}
\label{sec:separate}

Since cosmological dark matter (DM) production and annihilation is a local process, it is governed by the local properties of the radiation bath (e.g.\ temperature) and the local expansion rate, as well as the cross sections involved in any given model. As long as the spatial modulation due to the initial curvature fluctuation is on scales larger than the horizon, the modulation in the abundance $Y=n_\dm/s$ can be calculated locally using the so-called ``separate universe'' approach. Here $n_\dm$ is the dark matter number density and $s$ is the entropy density of the thermal bath.

In the separate universe approach, the local density perturbation $\delta_L$ associated with the long wavelength curvature fluctuation $\zeta$ is reabsorbed into the local Friedmann-Robertson-Walker parameters of the separate universe \cite{Sirko:2005uz,Li:2014sga}, explicitly constructed for radiation domination in a spatially flat background in Refs.~\cite{Zegeye:2021yml,Inomata:2023faq}, as we shall now review.

Specifically, observers in free-fall that are initially at rest with respect to the global expansion define the synchronous gauge, and in radiation domination, the density fluctuation in Fourier space is
\begin{equation}
\delta_L
= \frac{1}{3} \left(\frac{k}{aH}\right)^2 \zeta \propto a^2,
\label{eq:deltaL}
\end{equation}
when the comoving wavenumber is above the horizon $k\ll a H$, where the Hubble rate $H=d\ln a/dt$.
The local radiation density of the separate universe is then
\begin{equation}
\rho_L = \rho(1+\delta_L),
\end{equation}
which in turn defines the local scale factor as
\begin{equation}
 a_L = a(1-\delta_L/4)
 \label{eq:aL}
\end{equation}
at equal synchronous times.
Because dark matter production depends crucially on the local expansion rate, we need to know the local Hubble rate $H_L(a_L)$ in the presence of $\delta_L$. Taking the time derivative of Eq.~(\ref{eq:aL}), we can see that 
\begin{equation}
H_L^2(a_L(t)) = H^2(a(t))(1-\delta_L).
\label{eq:Hubbletime}
\end{equation}
The local Friedmann equation associates this change with both the local radiation density and the spatial curvature $K_L$ associated with the curvature fluctuation:
\begin{equation}
H_L^2(a_L) \equiv \frac{8\pi G \rho_L}{3} - \frac{K_L}{a_L^2}. 
\label{eq:Hubblescalefactor}
\end{equation}
By comparing Eqs.~(\ref{eq:Hubbletime}) and (\ref{eq:Hubblescalefactor}), the local spatial curvature induced by $\delta_L$ is identified as
\begin{equation}
\frac{K_L}{a_L^2} = 2 H^2 \delta_L,
\end{equation}
which is consistent with the perturbation to the 3D Ricci scalar induced by the curvature fluctuation \cite{Inomata:2023faq}. 
Notice that when compared to the global universe at the same value of the scale factor $a_L=a$,
$\rho_L(a_L) = \rho(a_L)$ since $T_L=T$
and the only change is that the expansion rate differs due to the spatial curvature since
\begin{equation}
H_L^2(a_L) = \frac{8\pi G \rho(a_L)}{3} (1- 2 \delta_L) = {H}^2(a_L) (1 - 2 \delta_L).
\end{equation}
Here and below, all variables are functions of $a_L$ unless otherwise specified.

Likewise in local coordinates, the local entropy is conserved $s_L(a_L)=s(a_L) \propto a_L^{-3}$ and the local temperature evolves in the usual way given changes in the relativistic degrees of freedom, so we can calculate the local DM abundance $Y_L$ by solving the usual Boltzmann system with all local quantities (e.g.~\cite{Binder:2017rgn})\footnote{In Ref.~\cite{Binder:2017rgn}, $\sv_\noteq$ is denoted as $\sv_{\rm neq}$ and the abundance is evolved as a function of temperature.},
\begin{equation}
\label{eq:boltzmann}
\frac{d Y_L}{d\ln a_L}= \frac{s_L }{H_L} \left[  \sv  Y_\eq^2 - \sv_\noteq  Y_L^2  \right],
\end{equation}
where $Y_\eq$ is the DM abundance if it were in local thermodynamic equilibrium with the radiation bath and $\sv$ is the thermally averaged DM annihilation cross section. $\sv_\noteq$ is the annihilation cross section similarly averaged over the actual, potentially non-equilibrium (``$\noteq$"), dark matter phase space distribution. We shall see in our examples below that the two cross sections are typically equal $\sv_\noteq \approx \sv$ when the dark matter is in kinetic equilibrium with the radiation and can be simply approximated otherwise. In this section we keep the methodology general and consider these cross sections to be arbitrary functions of the local scale factor.
 
Since the fluctuation in $Y_L$ due to $\delta_L$ is small, we can linearize Eq.~(\ref{eq:boltzmann}) in
\begin{equation}
\delta_Y(a_L)  \equiv \frac{Y_L(a_L) - {Y}(a_L)}{ Y(a_L)}
\end{equation}
and $\delta_L$ to obtain
\begin{align}
\frac{d \delta_Y}{d\ln a_L} ={}& \frac{s}{ H Y } [ \sv (\delta_L-\delta_Y) Y_\eq^2 \nonumber\\
{}&\quad -
\sv_\noteq( \delta_L+\delta_Y )  Y^2  ].
\label{eq:mainequation}
\end{align}
Here $s/HY = s(a_L)/H(a_L)Y(a_L)$ with their functional forms the same as in the global universe. 

Note that in the absence of the local curvature effect of $\delta_L$ on the expansion rate $H_L(a_L) \rightarrow H(a_L)$, $Y_L(a_L)\rightarrow Y(a_L)$, and $\delta_Y\rightarrow 0$ in spite of the finite local density fluctuation. Physically then $\delta_Y$ represents the fractional change in the abundance due to the effect of local curvature on the local expansion alone. Therefore the only way that $\delta_Y$ can be generated from the curvature fluctuation is through this $K_L$ effect. This includes any changes to the radiation temperature $T(a)$ due to entropy injection from particles annihilating away in the bath which changes the relationship between density fluctuations and temperature fluctuations and causes transient changes to the equation of state of the background (cf.\ Ref.~\cite{Bellomo:2022qbx} v2).

We refer to the effect of $\delta_L Y_\eq^2$ in Eq.~(\ref{eq:mainequation}) as modulated production and that of $\delta_L Y^2$ as modulated annihilation. For cases where modulated production is important, we generically expect that since the final abundance $Y_L(\infty)$ involves the competition between reaction rates and the Hubble rate, that the $\mathcal{O}({\delta_L})$ change in the Hubble rate at production will lead to 
\begin{equation}
\delta_Y(a_*) = \mathcal{O}(\delta_L(a_*)),
\label{eq:generationscaling}
\end{equation}
where $a_*$ is evaluated at the characteristic ``freezing" epoch where $Y_L$ is sufficiently close to its final value $Y_L(\infty)$. The coefficient and sign will depend on how a larger or smaller Hubble rate affects the abundance (see \S \ref{sec:results}). For definiteness we take $a_*$ to be defined by 
\begin{equation}
\Big| \ln \frac{Y_L(a_*)}{Y_L({\infty})}\Big|  = 1.
\label{eq:astar}
\end{equation} 

After $a_*$, the modulated abundance $\delta_Y$ can still change since $\delta_L$ itself grows. Since $Y\gg Y_\eq$ for non-relativistic dark matter, this mainly happens through  modulated annihilation where Eq.~(\ref{eq:mainequation}) can be approximated as
\begin{align}
\frac{d \delta_Y}{d\ln a_L} \approx {}&- \frac{s Y}{ H } 
\sv_\noteq \delta_L   =  -\frac{n_\dm}{ H } 
\sv_\noteq \delta_L   .
\end{align}
Here $Y$ is nearly constant, reflecting a small annihilation rate vs.~Hubble rate $n_\dm \sv_\noteq /H \ll 1$ but so long as this decreases more slowly than $\delta_L$ increases, $\delta_Y$ will continue to change. If we assume that for some range of time, not necessarily during radiation domination,
\begin{equation}
\sv_\noteq(a) \propto a^{p}
\label{eq:svscaling}
\end{equation}
the relevant comparison for whether late or early time annihilation is more important during this interval is whether
\begin{equation} \label{eq:mod_annihil}
\frac{ n_\dm  \sv_\noteq }{ H} \delta_L \propto  a^{p+ (9w-1)/2}
\end{equation}
grows or decays, where $w=P/\rho$ and $\delta_L\propto (k/aH)^2$. Note that here and below, we do not distinguish evaluation at $a$ vs.~$a_L$ for quantities that are already first order in $\delta_L$. 

If $p > (1-9w)/2$ then the modulated annihilation will be dominated in this interval by late times and early times otherwise. For radiation domination when $w=1/3$, the transition is for $p=-1$, and for matter domination where $w=0$ it is for $p=1/2$. Therefore for $-1< p < 1/2$, the dominant annihilation modulation will occur around matter-radiation equality $a_\mr$ and we can solve for $\delta_Y$, assuming again that $n_\dm \sv_\noteq /H \ll 1$ and modulated annihilation dominates over production, to obtain
\begin{equation}
\delta_Y(a) \approx -\frac{f_\mr}{p+1}
\delta_L(a_\mr) \times \begin{cases}
 ({a}/{a_\mr})^{p+1}  & a \lesssim a_{\mr} \\
 {\cal O}(1) & a > a_{\mr}
 \end{cases},
 \label{eq:annihilationscaling}
\end{equation}
where the ${\cal O}(1)$ accounts for the transition to matter domination when modulated annihilation ceases. Here the normalization constant $f_\mr$ gives the ratio of annihilation to Hubble rates at equality
\begin{equation}
f_{\mr}=\sqrt{2} \frac{n_{\dm} \sv_\noteq }{H} \Big|_{a_\mr},
\end{equation}
where $\sqrt{2}$ accounts for the fact that at equality the radiation contributes half the total energy density so that during radiation domination $ n_\dm \sv_\noteq/H = f_\mr (a/a_\mr)^{p-1}$.

A final case that will be relevant is when $p$ transitions from $p>-1$ to a value $<-1$ during radiation domination, say at an epoch $a_\kd < a_\mr$. Then the modulated annihilation contribution is given by 
\begin{equation}
\delta_Y(a) \approx -\frac{f_\kd}{p+1}
\delta_L(a_\kd) \times \begin{cases}
 ({a}/{a_\kd})^{p+1}  & a \lesssim a_{\kd} \\
 {\cal O}(1) & a > a_{\kd}
 \end{cases},
 \label{eq:kineticdecouplingscaling}
\end{equation}
where the normalization constant 
\begin{equation}
f_{\kd}=\frac{n_{\dm} \sv_\noteq }{H} \Big|_{a_\kd}
\end{equation}
and the ${\cal O}(1)$ coefficient depends on how smoothly the transition occurs.

In both the $a_{\mr}$ and $a_\kd$ dominated annihilation cases, $p=-1$ is a special case where $|\delta_Y|$ grows logarithmically with $a_L$ up to that epoch before freezing in. In those cases modulo the $1/(1+p)$ factors, Eq.~(\ref{eq:annihilationscaling}) and Eq.~(\ref{eq:kineticdecouplingscaling}) give the order of magnitude up to log factors rather than the exact value of the result for $a>a_{\mr,\kd}$ and so we will consider this special case as part of those generic scalings.

We can combine these generic expectations for the contributions to $\delta_Y$ at $a_*$ and annihilation at $a\gg a_*$ in Eqs.~(\ref{eq:generationscaling}), (\ref{eq:annihilationscaling}), and (\ref{eq:kineticdecouplingscaling}) to determine when each dominates. In either the $a_{\mr}$ and $a_{\kd}$ cases, modulated annihilation dominates over production in determining $\delta_Y$ if
\begin{equation}
f_{\kd,\mr} \left( \frac{a_{\kd,\mr}}{a_*} \right)^2 \gg 1,
\label{eq:prodvsanih}
\end{equation}
i.e.\ if the epochs of annihilation and production are separated by a large enough factor for the growth of $\delta_L$ to overcome the small ratio of annihilation to Hubble rates.

Finally it is interesting to note that there is in principle a difference between $\delta_Y$, the change in the abundance when the local universe reaches the same scale factor or temperature as the global universe, and the spatial fluctuation of the abundance at the same synchronous time. For cases where the modulations are dominated by early times, at late times $\delta_Y$ itself is frozen and the difference in evaluation time between the two quantities is irrelevant. In this case the isocurvature mode is time slicing invariant. More generally, in radiation domination where 
\begin{equation}
\ln a_L - \ln a = -\delta_L/4
\end{equation}
the synchronous gauge dark matter photon isocurvature perturbation at fixed time well after electron-positron annihilation is
\begin{align}
S & \equiv \frac{\delta(n_\dm/n_\gamma)}{n_\dm/n_\gamma}
=
\frac{ Y_L(a_L(t))- Y(a(t))}{ Y(a(t))} \nonumber\\
& = \delta_Y - \frac{d \ln  Y}{d\ln a}\frac{{\delta_L}}{4}, 
\qquad (a\ll a_\mr),
\end{align}
where the last expression can be evaluated at $t$ to linear order in $\delta_L$. The second term accounts for the gauge dependence of adiabatic abundance fluctuations defined from a time evolving abundance and in that sense does not reflect a true isocurvature perturbation on its own. Furthermore note that even for cases like $\sv_\noteq=\,$const.\ and more generally, the $p<1/2$ modulated annihilation cases of Eq.\ (\ref{eq:annihilationscaling}), $S\rightarrow \delta_Y$ when $a \gg a_\mr$. For cases that are dominated by modulated production (see Eq.~\ref{eq:generationscaling}) or modulated annihilation at $a_\kd$ (see Eq.~\ref{eq:kineticdecouplingscaling}), this convergence occurs even earlier. Because of this late time equivalence, we use $\delta_Y$ in all cases as the measure of isocurvature generation.

\section{Freeze-in/out models}
\label{sec:models}

\begin{figure}[ht]
    \centering
    \includegraphics[width=0.5\linewidth]{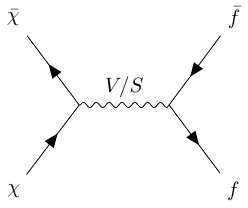}
    \caption{Feynman diagram for $\bar{\chi} \chi \to \bar{f} f$ annihilation.}
    \label{fig:feynmandiagramVS}
\end{figure}

To illustrate the range of possible phenomena, we consider two generic dark matter particle models in which a fermionic dark matter particle, $\chi$ interacts with standard model fermions, $f$, via a vector ($V$) or scalar ($S$) mediator, as described in \cite{Berlin:2014tja}. The Feynman diagram for DM annihilation is shown in Fig.~\ref{fig:feynmandiagramVS}.

The interaction Lagrangian for the vector-mediated scenario is:
\begin{equation}
    \mathcal{L} \supset g_\chi \bar{\chi} \gamma^\mu V_\mu \chi + g_f \bar{f} \gamma^\mu V_\mu f,
    \label{eq:LVmed}
\end{equation}
where $g_\chi$ and $g_f$ represent the coupling strength of the dark matter and the SM fermions to $V$, respectively and the particles $\chi$ and $V$ have masses $m_\chi$ and $m_V$. An example of a specific realization of this broader model class is the millicharged dark matter, used as the example in \cite{Bellomo:2022qbx}, in which the millicharge arises from kinetic mixing with a dark photon mediator to the SM, or from a DM hypercharge.

From the above diagrams, one can compute the cross section for $\chi \bar{\chi}$ annihilation from Eq.~\eqref{eq:LVmed}:
\begin{align}
    \sigma_{\bar{\chi}\chi \to \bar{f}f} &= \frac{g_f^2 g_\chi^2}{12 \pi  s \left[\left(s-m_V^2\right)^2 + m_V^2 \Gamma_V^2 \right]}  \nonumber \\
    &\times \sqrt{\frac{s-4 m_f^2}{s-4 m_\chi^2}} \left(s+2 m_f^2\right) \left(s+2 m_\chi^2\right),
    \label{eq:sigmaVmed}
\end{align}
where $s$ is the Mandelstam variable for the center-of-mass energy squared (not to be confused with entropy $s$ elsewhere), and $\Gamma_V$ is the total decay width of $V$, which can be neglected away from resonance when $m_\dm \gg m_V$.

The interaction Lagrangian for the scalar-mediated scenario is:
\begin{equation}
    \mathcal{L} \supset \lambda_\chi \bar{\chi} \chi S + \lambda_f \bar{f} f S,
    \label{eq:LSmed}
\end{equation}
where $\lambda_\chi$ and $\lambda_f$ are the couplings from $S$ to $\chi$ and $f$ respectively. In this case, the annihilation cross section is:
\begin{align}
    \sigma_{\bar{\chi}\chi \to \bar{f}f} &= \frac{\lambda_f^2 \lambda_\chi^2 }{16 \pi s \left[\left(s - m_S^2\right)^2 + m_S^2 \Gamma_S^2\right]} \nonumber \\
    &\times \sqrt{\frac{s - 4 m_f^2}{s - 4 m_\chi^2}} (s - 4 m_f^2) (s - 4 m_\chi^2),
    \label{eq:sigmaSmed}
\end{align}
where $m_S$ is the mass of the scalar mediator and we again neglect its decay width, $\Gamma_S$. Note that there can also be axial vector couplings ($\sim \gamma^5$) for each of these interactions, which can in principle change the velocity and temperature dependence of the cross section. For simplicity, we set these to zero, but our technique itself can be used to predict isocurvature production given any such cross section, or more generally whenever the abundance itself can be calculated from local quantities.

For different choices of parameter values (couplings $g_\chi$, $g_f$, $\lambda_\chi$, $\lambda_f$ and masses $m_\chi$, $m_V$, $m_S$), these models can lead to both freeze-in or freeze-out DM production. We construct four illustrative examples, a freeze-in and freeze-out scenario for each of the two models. For simplicity, we consider that the mediator interacts only with the DM and electrons ($f=e$). We take $m_\chi = 100\,\mathrm{GeV}$ in all cases and avoid resonances (which introduce more complicated temperature dependence) by choosing the mediator masses to be $m_V = m_S = 1\,\mathrm{GeV}$. The coupling parameters are then chosen to produce the observed DM abundance
\begin{equation}
Y(\infty) \approx  
\frac{\rho_{\chi}}{{m_\chi s}}\Big|_{a=1}  \approx 3.64\,  \Omega_\dm h^2 \left( \frac{\rm eV}{m_\dm}\right)
\end{equation}
with
$s \approx 7.04 n_\gamma(T_{\rm CMB})$ and
$\Omega_\dm h^2=0.12$ \cite{Planck:2018vyg}.
For freeze-out, this requires $g_\chi = g_e = 0.1774$ and $\lambda_\chi = \lambda_e = 0.5044$ and for freeze-in, it requires $g_\chi = g_e = 2.716 \times 10^{-6}$ and $\lambda_\chi = \lambda_e = 3.259 \times 10^{-6}$. We do not claim that these illustrative cases are fully viable DM models given direct and indirect detection constraints (see e.g.~\cite{Workman:2022ynf}). Rather, our aim is to illustrate the mechanisms that generate isocurvature perturbations in a range of simple, but representative scenarios.

For a DM annihilation cross section $\sigma$, assuming the DM phase space is distributed according to Maxwell-Boltzmann statistics with a temperature $T$, the thermally averaged cross section is \cite{GondoloCosmicAbundancesStable1991}
\begin{equation} \label{eq:thermalaverage}
    \langle\sigma v\rangle = \frac{1}{8m^4 T K_2^2\!\left(\frac{m}{T}\right)} \int_{4m^2}^{\infty} ds \, \sigma  \sqrt{s} \left(s - 4m^2\right) K_1\!\left(\tfrac{\sqrt{s}}{T}\right)
\end{equation}
where $m$ is the DM mass, $s$ is the squared center-of-mass energy in the collisions, and $K_n$ is the modified Bessel function of the second kind of order $n$. The main practical difference between the vector and scalar-mediated DM models is the low-temperature/velocity dependence in the thermally averaged cross section. The asymptotic temperature dependence of both models can easily be found from Eqs.~\eqref{eq:sigmaVmed} and~\eqref{eq:sigmaSmed} and is summarized in Tab.~\ref{tab:sigmav}. These scalings originate from the fact that the vector-mediated DM has $s$-wave-suppressed annihilation while the scalar-mediated model is $p$-wave-suppressed. 

{
\setlength{\tabcolsep}{10pt}
\renewcommand{\arraystretch}{1.5}
\begin{table}[h]
\centering
\begin{tabular}{|c|c|c|} \hline
\textbf{Model} & \textbf{High} $\bm{T}$ & \textbf{Low} $\bm{T}$ \\ \hline
Vector-mediated & $\sv \propto T^{-2}$ & $\sv \propto \mathrm{const}$ \\ 
Scalar-mediated & $\sv \propto T^{-2}$ & $\sv \propto T$ \\ \hline
\end{tabular}
\caption{Asymptotic temperature dependence of the thermally averaged cross section for each model.}
\label{tab:sigmav}
\end{table}
}

We apply these thermally averaged cross sections to Eq.~(\ref{eq:mainequation}) by taking for the production term
\begin{equation}
\sv(a) = \sv(a(T))
\end{equation}
with the thermal bath temperature given by conservation of entropy through $s(a)  = (2\pi^2/45)\, g_{\star,s} T^3 \propto a^{-3}$, where $g_{\star,s}$ is the usual effective number of relativistic degrees of freedom in entropy \cite{Laine:2015kra,Borsanyi:2016ksw,Husdal:2016haj}.

For the annihilation term, kinetic decoupling affects the evolution of $\sv_\noteq(a)$. A DM particle undergoing freeze-out is initially both in chemical and kinetic equilibrium, while for freeze-in, it is never in chemical equilibrium, and possibly never in kinetic equilibrium. In both cases, even after number-changing production and annihilation reactions stop, there are typically also scattering reactions that may keep $T_\dm=T$ via momentum transfer. In practice, for freeze-out models, we assume that the DM has the same temperature as the radiation bath until momentum transfer effectively ceases and it kinetically decouples \cite{Binder:2017rgn}:
\begin{equation}
\sv_\noteq(a) = \sv(a(T_\dm)) 
\end{equation}
with
\begin{equation}
    T_\dm(a) \approx \begin{cases}
        T & T > T_\kd \\
        T_\kd \, (a / a_\kd)^{-2} & T < T_\kd
    \end{cases}.
\end{equation}
The scaling for $T<T_\kd$ comes from the fact that after kinetic decoupling, particle momenta always redshift as $a^{-1}$ and the dark matter is assumed to be non-relativistic. Notice that in terms of our scaling solutions in the previous section where in Eq.~(\ref{eq:svscaling}) we parameterized $\sv_\noteq \propto a^p$, this scaling is equivalent to $\sv_\noteq \propto T_\dm^{-p}$ before kinetic decoupling (except when $g_{\star,s}$ changes) and $\sv_\noteq \propto T_\dm^{-p/2}$ after, when the dark matter is non-relativistic. Notice that for the scalar-mediated case this makes the low temperature scaling change from $p=-1$ to $p=-2$ across $a_\kd$ (cf.~Eq.~\ref{eq:kineticdecouplingscaling}). For illustrative purposes, we simply set the temperature of dark matter kinetic decoupling to be $T_\mathrm{kd} = 10^{-4}  m_\dm$ \cite{Bringmann:2009vf} but it should be considered an adjustable quantity that can be calculated in a given model from the momentum exchange rate. The accuracy of this sort of prescription for $\sv_\noteq$ has been tested in Refs.~\cite{Binder:2017rgn,DuRevisitingDarkMatter2022}.

For freeze-in models, we are typically in a region of parameter space where annihilation is never important which we check by monitoring the contribution of the maximal annihilation case where $\sv_\noteq(a) = \sv(a)$ and testing Eq.~(\ref{eq:prodvsanih}).

In each case, we illustrate the generation of isocurvature from curvature with a $k$-mode that is near the horizon at matter radiation equality: $k_{\mr}= a_{\mr} H_{\mr}$. For this mode when $a\lesssim a_{\mr}$,
\begin{equation}
\frac{\delta_L}{\zeta} = \frac{2}{3} \left(\frac{a}{a_\mr}\right)^2,
\label{eq:mode}
\end{equation}
and we correspondingly compute $\delta_Y/\zeta$, which is the quantity that is observationally constrained (see \S\ref{sec:discussion}). Other modes that are superhorizon scaled during the relevant modulation epoch simply differ in the normalization through Eq.~(\ref{eq:deltaL}).

For simplicity we ignore the matter contribution to the expansion rate as well as any annihilation products, even though we compute up to $a=a_\mr$, by taking $\rho = (\pi^2/15)\, g_\star T^4$, where $g_\star$ is the effective number of relativistic degrees of freedom in energy density. For each of these models, we first numerically solve Eq.~(\ref{eq:boltzmann}) for the background and then solve Eq.~(\ref{eq:mainequation}) for $\delta_Y$ with $\delta_L$ normalized for the mode $k_\mr$ in Eq.~(\ref{eq:mode}).

\section{Model Results}
\label{sec:results}

\begin{figure}[tbp]
    \centering
    \includegraphics[width=\columnwidth]{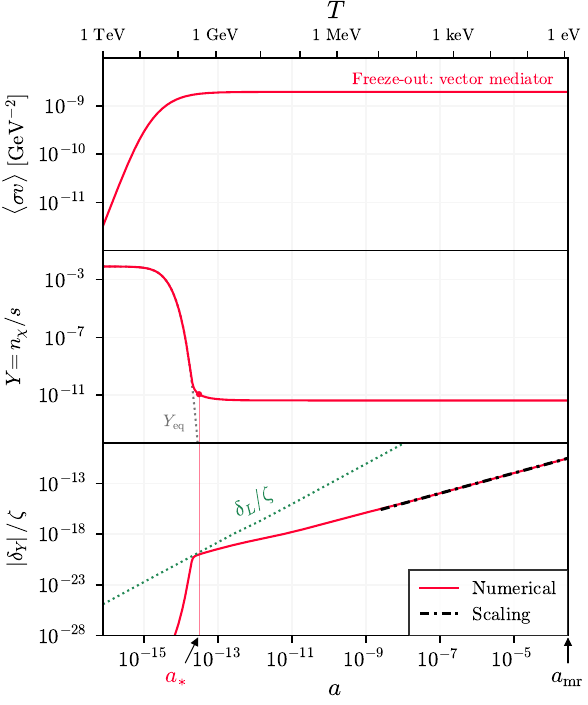}
    \caption{Isocurvature production for the vector mediated freeze-out case  where modulated annihilation around matter-radiation equality dominates. Top: thermally averaged annihilation cross section $\sv$, which reaches a constant at low $T$ or large scale factor $a$. Middle: dark matter abundance $Y$ in the background, which follows the equilibrium abundance $Y_{\rm eq}$ before freezing out at $a_*$. Bottom: isocurvature vs.~curvature fraction $\delta_Y/\zeta$ (negative definite) for the horizon wavenumber at matter-radiation equality $a_\mr$, which grows from a freeze-out value ${\cal O}(\delta_L(a_*)/\zeta)$ (dotted line) linearly in $a$ until $a_\mr$ matching the scaling solution (dot-dashed line, Eq.~(\ref{eq:annihilationscaling})). } 
    \label{fig:vectorFO}
\end{figure}

\begin{figure}[tbp]
    \centering
    \includegraphics[width=\columnwidth]{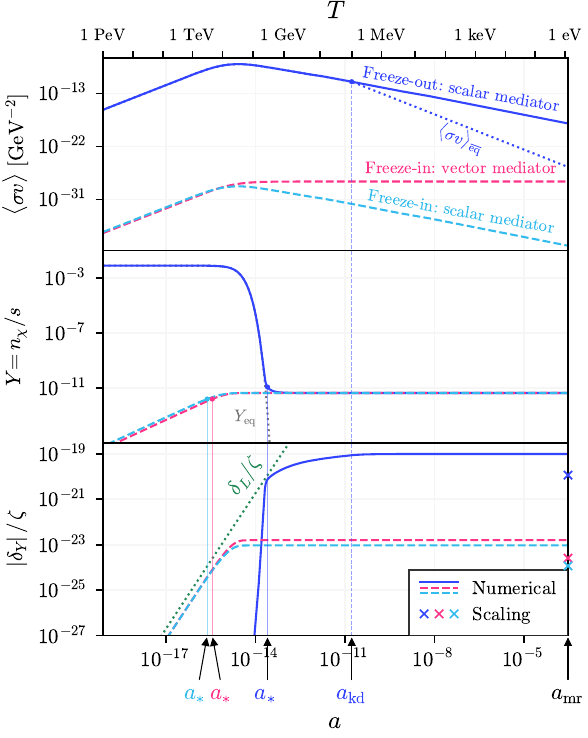}
    \caption{Isocurvature production for cases where modulation occurs before matter radiation equality (freeze-out scalar mediator [solid line], freeze-in scalar and vector mediator [dashed lines]). Labels are otherwise as in Fig.~\ref{fig:vectorFO}. For the freeze-in cases, production ceases around $a_*$ with the final $\delta_Y/\zeta= {\cal O}(\delta_L(a_*)/\zeta)$ (positive definite). For the freeze-out scalar mediated case, modulated annihilation grows logarithmically until kinetic decoupling $a_{\kd}$ and freezes into a negative definite value thereafter. The expectations from the scaling relations (\ref{eq:generationscaling}) and (\ref{eq:kineticdecouplingscaling}) are shown as $\mathsf{x}$ markers and are valid to order of magnitude.}
    \label{fig:scalarFOFI}
\end{figure}

In this section, we present the results for isocurvature generation from curvature fluctuations for the illustrative models described in \S \ref{sec:models}. We start with the vector-mediated freeze-out model in Fig.~\ref{fig:vectorFO}. Here $\sv$ approaches a constant at low velocities and temperatures, as shown in the top panel, so that kinetic decoupling does not change annihilation at late times. In this case freeze-out as defined by Eq.~(\ref{eq:astar}) occurs at $a_* = 3.1 \times 10^{-14}$ as shown in the middle panel. At this epoch $\delta_Y(a_*) = -0.58\, \delta_L(a_*)$ validating the expectations of Eq.~(\ref{eq:generationscaling}). The sign of the modulation is negative since $\delta_L>0$ decreases the local Hubble rate and allows annihilation to continue locally later than the in the background. On the other hand, since the cross section reaches a constant ($p=0$), modulated annihilation continues after $a_*$. In fact the analytic scaling solution from Eq.~(\ref{eq:annihilationscaling}) $\delta_Y(a)= -(a/a_\mr) f_\mr \delta_L(a_{\mr})$ agrees very closely with the numerical calculation in this regime as also shown in the dashed black line in Fig.~\ref{fig:vectorFO}. Therefore somewhat counterintuitively, the main effect of modulated freeze-out in this case occurs much much later than the nominal freeze-out epoch $a_*$ due to the growth of density perturbations outside the horizon. Since the mode considered is on the horizon at matter-radiation equality and would not satisfy the separate-universe assumption afterward, we do not continue the calculation to $a>a_\mr$, though we have shown through the scaling relation of Eq.~(\ref{eq:annihilationscaling}) that for this case, $\delta_Y(\infty) \sim -f_\mr \delta_L(a_{\mr})$, asymptoting to a constant on superhorizon scales.

We contrast this with the 3 other cases where $\delta_Y$ approaches a constant well before matter-radiation equality as shown in Fig.~\ref{fig:scalarFOFI}. In this class, we first consider the scalar-mediated freeze-out case. Here the annihilation cross section is velocity-dependent at low temperatures and kinetic decoupling affects how quickly $\sv_\noteq$ decreases and thus when modulated annihilation stops. Here we take kinetic decoupling to occur at $T=10\,\mathrm{MeV}$ or $a_\kd = 1.7 \times 10^{-11}$. Notice that at $a<a_\kd$, $p=-1$ and $|\delta_Y|$ grows logarithmically. Once $a>a_\kd$, $p=-2$ and modulated annihilation freezes out, here to a constant level $\delta_Y(\infty) = -8.5 f_\kd\delta_L(a_\kd)$ validating the expected scaling (i.e.~Eq.~(\ref{eq:kineticdecouplingscaling}) without the $(1+p)^{-1}$ factor).

Finally we consider the vector and scalar freeze-in cases. In both cases, the modulated annihilation term is irrelevant as is whether the produced dark matter is kinetically decoupled. The whole effect therefore comes from modulated production and the isocurvature mode reaches a constant $\delta_Y(\infty) = 6.2\,\delta_L(a_*)$ for vector mediated and $8.0\,\delta_L(a_*)$ for scalar mediated cases. Notice first that the sign of $\delta_Y/\delta_L$ is positive and opposite to that of freeze-out because a smaller local Hubble rate means that production continues to occur at a later time than in the background. Also, since the cross section decreases at low temperature in the scalar case, to match the same final abundance $a_*$ and hence $\delta_Y$ itself is smaller. Otherwise the two freeze-in cases behave very similarly despite the difference in the temperature scaling of the cross section, unlike the corresponding freeze-out cases.

\section{Discussion}
\label{sec:discussion}

Our separate universe methodology provides a simple means of determining the isocurvature production from a curvature fluctuation $\zeta$ for any model where the cosmological expansion rate affects the final dark matter abundance, including generic freeze-out and freeze-in cases. One merely needs to recalculate the abundance using the local rather than global background. In this method, it is the change due to $\zeta$ of the local expansion rate from the space curvature contribution to the Friedmann equation that modulates the abundance, not the change to the local density of the plasma, though we use this fluctuation $\delta_L$ to parameterize the results. 

For a typical model, there are three important epochs in this process: $a_*$ when the freezing of the background abundance occurs, $a_\kd$ when the annihilation rate can change relative to the Hubble rate due to kinetic decoupling of the dark matter, and $a_\mr$ when matter dominates the Hubble rate and changes its scaling relative to the annihilation rate. In typical freeze-in models, annihilation is never important and so the isocurvature amplitude scales as $\delta_Y \sim \delta_L(a_*)$, and is fully correlated with the curvature $\zeta$. For freeze-out models, modulated annihilation can occur at $a\gg a_*$ due to the growth of density perturbations. For cases where the annihilation rate drops sufficiently due to kinetic decoupling this leaves $\delta_Y \sim - f(a_\kd) \delta_L(a_\kd)$ which is fully anti-correlated with $\zeta$, where the annihilation to Hubble rate $f$ is evaluated at $a_\kd$. For cases like a constant thermally averaged annihilation cross section, modulated annihilation continues to grow until $a_\mr$ and the anticorrelation scales as $\delta_Y \sim - f(a_\mr) \delta_L(a_\mr)$. In fact it is these cases where the isocurvature generation is largest and its size increases with increasing annihilation rate $f(a_\mr)\propto m_{\dm}^{-1}$ for a fixed relic mass density.

In all of these cases, which we have illustrated using concrete scalar and vector mediated dark matter models, the relative amplitude of the correlated or anticorrelated isocurvature mode to the curvature mode is highly suppressed on scales relevant for the CMB as long as $a_* \ll a_{\mr}$ due to the either the smallness of the superhorizon density perturbation at that time or the smallness of the annihilation rate after that time. A generic model therefore will easily evade CMB bounds from Planck which constrain a primordial $\delta_Y=S$ from inflation to $(S/\zeta)^2<10^{-3}$ at 95\% C.L. \cite{Planck:2018jri} for fully correlated fluctuations of the same spectrum. This case behaves similarly to our freeze-in and scalar mediated freeze-out in that the isocurvature fluctuation is frozen in at early times but has a different $k$-spectrum due to the fact that $\delta_L \propto (k/aH)^2\zeta$. In principle constraints on such a spectrum place an independent bound on how close $a_*$ can be to $a_\mr$ and how much annihilation can occur near recombination but other more direct bounds on these quantities are generally much stronger in typical models (e.g.~\cite{Kawasaki:2021etm}). 

For the freeze-out vector mediated case, the isocurvature modes themselves grow until matter radiation equality and would impact observable CMB modes in a different way that is beyond the scope of the separate universe approximation since the relevant modes are subhorizon at recombination and annihilation products can change the ionization. Even in this case, we can infer from the smallness of $\delta_Y/\zeta$ that such models generically predict signals that are well below cosmic variance limits for the CMB. 

More generally, our methodology provides a model-independent means of determining the generation of isocurvature from curvature that is as simple as determining the background dark matter abundance itself and moreover illuminates the amplitude and the relevant scales in this process. 

\smallskip

\textit{Note added after completion:} Some aspects of this work are mirrored in the independent work of \cite{Stebbins}.

\acknowledgments

We thank Dan Hooper, Austin Joyce, and Gordan Krnjaic for useful conversations. I.H. was supported by a generous contribution from Philip Rice. W.H. was supported by U.S. Dept.\ of Energy contract DE-FG02-13ER41958 and the Simons Foundation. L.J.\ was supported by the Kavli Institute for Cosmological Physics at the University of Chicago through an endowment from the Kavli Foundation and its founder Fred Kavli.

\bibliographystyle{apsrev4-2}
\bibliography{ref}

\begin{thebibliography}{31}%
\makeatletter
\providecommand \@ifxundefined [1]{%
 \@ifx{#1\undefined}
}%
\providecommand \@ifnum [1]{%
 \ifnum #1\expandafter \@firstoftwo
 \else \expandafter \@secondoftwo
 \fi
}%
\providecommand \@ifx [1]{%
 \ifx #1\expandafter \@firstoftwo
 \else \expandafter \@secondoftwo
 \fi
}%
\providecommand \natexlab [1]{#1}%
\providecommand \enquote  [1]{``#1''}%
\providecommand \bibnamefont  [1]{#1}%
\providecommand \bibfnamefont [1]{#1}%
\providecommand \citenamefont [1]{#1}%
\providecommand \href@noop [0]{\@secondoftwo}%
\providecommand \href [0]{\begingroup \@sanitize@url \@href}%
\providecommand \@href[1]{\@@startlink{#1}\@@href}%
\providecommand \@@href[1]{\endgroup#1\@@endlink}%
\providecommand \@sanitize@url [0]{\catcode `\\12\catcode `\$12\catcode `\&12\catcode `\#12\catcode `\^12\catcode `\_12\catcode `\%12\relax}%
\providecommand \@@startlink[1]{}%
\providecommand \@@endlink[0]{}%
\providecommand \url  [0]{\begingroup\@sanitize@url \@url }%
\providecommand \@url [1]{\endgroup\@href {#1}{\urlprefix }}%
\providecommand \urlprefix  [0]{URL }%
\providecommand \Eprint [0]{\href }%
\providecommand \doibase [0]{https://doi.org/}%
\providecommand \selectlanguage [0]{\@gobble}%
\providecommand \bibinfo  [0]{\@secondoftwo}%
\providecommand \bibfield  [0]{\@secondoftwo}%
\providecommand \translation [1]{[#1]}%
\providecommand \BibitemOpen [0]{}%
\providecommand \bibitemStop [0]{}%
\providecommand \bibitemNoStop [0]{.\EOS\space}%
\providecommand \EOS [0]{\spacefactor3000\relax}%
\providecommand \BibitemShut  [1]{\csname bibitem#1\endcsname}%
\let\auto@bib@innerbib\@empty
\bibitem [{\citenamefont {Wands}\ \emph {et~al.}(2000)\citenamefont {Wands}, \citenamefont {Malik}, \citenamefont {Lyth},\ and\ \citenamefont {Liddle}}]{Wands:2000dp}%
  \BibitemOpen
  \bibfield  {author} {\bibinfo {author} {\bibfnamefont {D.}~\bibnamefont {Wands}}, \bibinfo {author} {\bibfnamefont {K.~A.}\ \bibnamefont {Malik}}, \bibinfo {author} {\bibfnamefont {D.~H.}\ \bibnamefont {Lyth}},\ and\ \bibinfo {author} {\bibfnamefont {A.~R.}\ \bibnamefont {Liddle}},\ }\href {https://doi.org/10.1103/PhysRevD.62.043527} {\bibfield  {journal} {\bibinfo  {journal} {Phys. Rev. D}\ }\textbf {\bibinfo {volume} {62}},\ \bibinfo {pages} {043527} (\bibinfo {year} {2000})},\ \Eprint {https://arxiv.org/abs/astro-ph/0003278} {arXiv:astro-ph/0003278} \BibitemShut {NoStop}%
\bibitem [{\citenamefont {Weinberg}(2004)}]{Weinberg:2004kr}%
  \BibitemOpen
  \bibfield  {author} {\bibinfo {author} {\bibfnamefont {S.}~\bibnamefont {Weinberg}},\ }\href {https://doi.org/10.1103/PhysRevD.70.043541} {\bibfield  {journal} {\bibinfo  {journal} {Phys. Rev. D}\ }\textbf {\bibinfo {volume} {70}},\ \bibinfo {pages} {043541} (\bibinfo {year} {2004})},\ \Eprint {https://arxiv.org/abs/astro-ph/0401313} {arXiv:astro-ph/0401313} \BibitemShut {NoStop}%
\bibitem [{\citenamefont {Wands}(2008)}]{Wands:2007bd}%
  \BibitemOpen
  \bibfield  {author} {\bibinfo {author} {\bibfnamefont {D.}~\bibnamefont {Wands}},\ }\href {https://doi.org/10.1007/978-3-540-74353-8_8} {\bibfield  {journal} {\bibinfo  {journal} {Lect. Notes Phys.}\ }\textbf {\bibinfo {volume} {738}},\ \bibinfo {pages} {275} (\bibinfo {year} {2008})},\ \Eprint {https://arxiv.org/abs/astro-ph/0702187} {arXiv:astro-ph/0702187} \BibitemShut {NoStop}%
\bibitem [{\citenamefont {Akrami}\ \emph {et~al.}(2020)\citenamefont {Akrami} \emph {et~al.}}]{Planck:2018jri}%
  \BibitemOpen
  \bibfield  {author} {\bibinfo {author} {\bibfnamefont {Y.}~\bibnamefont {Akrami}} \emph {et~al.} (\bibinfo {collaboration} {Planck}),\ }\href {https://doi.org/10.1051/0004-6361/201833887} {\bibfield  {journal} {\bibinfo  {journal} {Astron. Astrophys.}\ }\textbf {\bibinfo {volume} {641}},\ \bibinfo {pages} {A10} (\bibinfo {year} {2020})},\ \Eprint {https://arxiv.org/abs/1807.06211} {arXiv:1807.06211 [astro-ph.CO]} \BibitemShut {NoStop}%
\bibitem [{\citenamefont {Bellomo}\ \emph {et~al.}(2023)\citenamefont {Bellomo}, \citenamefont {Berghaus},\ and\ \citenamefont {Boddy}}]{Bellomo:2022qbx}%
  \BibitemOpen
  \bibfield  {author} {\bibinfo {author} {\bibfnamefont {N.}~\bibnamefont {Bellomo}}, \bibinfo {author} {\bibfnamefont {K.~V.}\ \bibnamefont {Berghaus}},\ and\ \bibinfo {author} {\bibfnamefont {K.~K.}\ \bibnamefont {Boddy}},\ }\href {https://doi.org/10.1088/1475-7516/2023/11/024} {\bibfield  {journal} {\bibinfo  {journal} {JCAP}\ }\textbf {\bibinfo {volume} {11}},\ \bibinfo {pages} {024}},\ \Eprint {https://arxiv.org/abs/2210.15691} {arXiv:2210.15691 [astro-ph.CO]} \BibitemShut {NoStop}%
\bibitem [{\citenamefont {Rubin}\ and\ \citenamefont {Ford}(1970)}]{Rubin:1970zza}%
  \BibitemOpen
  \bibfield  {author} {\bibinfo {author} {\bibfnamefont {V.~C.}\ \bibnamefont {Rubin}}\ and\ \bibinfo {author} {\bibfnamefont {W.~K.}\ \bibnamefont {Ford}, \bibfnamefont {Jr.}},\ }\href {https://doi.org/10.1086/150317} {\bibfield  {journal} {\bibinfo  {journal} {Astrophys. J.}\ }\textbf {\bibinfo {volume} {159}},\ \bibinfo {pages} {379} (\bibinfo {year} {1970})}\BibitemShut {NoStop}%
\bibitem [{\citenamefont {Komatsu}\ \emph {et~al.}(2009)\citenamefont {Komatsu} \emph {et~al.}}]{WMAP:2008lyn}%
  \BibitemOpen
  \bibfield  {author} {\bibinfo {author} {\bibfnamefont {E.}~\bibnamefont {Komatsu}} \emph {et~al.} (\bibinfo {collaboration} {WMAP}),\ }\href {https://doi.org/10.1088/0067-0049/180/2/330} {\bibfield  {journal} {\bibinfo  {journal} {Astrophys. J. Suppl.}\ }\textbf {\bibinfo {volume} {180}},\ \bibinfo {pages} {330} (\bibinfo {year} {2009})},\ \Eprint {https://arxiv.org/abs/0803.0547} {arXiv:0803.0547 [astro-ph]} \BibitemShut {NoStop}%
\bibitem [{\citenamefont {Aghanim}\ \emph {et~al.}(2020)\citenamefont {Aghanim} \emph {et~al.}}]{Planck:2018vyg}%
  \BibitemOpen
  \bibfield  {author} {\bibinfo {author} {\bibfnamefont {N.}~\bibnamefont {Aghanim}} \emph {et~al.} (\bibinfo {collaboration} {Planck}),\ }\href {https://doi.org/10.1051/0004-6361/201833910} {\bibfield  {journal} {\bibinfo  {journal} {Astron. Astrophys.}\ }\textbf {\bibinfo {volume} {641}},\ \bibinfo {pages} {A6} (\bibinfo {year} {2020})},\ \bibinfo {note} {\href{https://doi.org/10.1051/0004-6361/201833910e}{[Erratum: Astron.Astrophys. 652, C4 (2021)]}},\ \Eprint {https://arxiv.org/abs/1807.06209} {arXiv:1807.06209 [astro-ph.CO]} \BibitemShut {NoStop}%
\bibitem [{\citenamefont {McDonald}(2002)}]{McDonald:2001vt}%
  \BibitemOpen
  \bibfield  {author} {\bibinfo {author} {\bibfnamefont {J.}~\bibnamefont {McDonald}},\ }\href {https://doi.org/10.1103/PhysRevLett.88.091304} {\bibfield  {journal} {\bibinfo  {journal} {Phys. Rev. Lett.}\ }\textbf {\bibinfo {volume} {88}},\ \bibinfo {pages} {091304} (\bibinfo {year} {2002})},\ \Eprint {https://arxiv.org/abs/hep-ph/0106249} {arXiv:hep-ph/0106249} \BibitemShut {NoStop}%
\bibitem [{\citenamefont {Bernal}\ \emph {et~al.}(2017)\citenamefont {Bernal}, \citenamefont {Heikinheimo}, \citenamefont {Tenkanen}, \citenamefont {Tuominen},\ and\ \citenamefont {Vaskonen}}]{Bernal:2017kxu}%
  \BibitemOpen
  \bibfield  {author} {\bibinfo {author} {\bibfnamefont {N.}~\bibnamefont {Bernal}}, \bibinfo {author} {\bibfnamefont {M.}~\bibnamefont {Heikinheimo}}, \bibinfo {author} {\bibfnamefont {T.}~\bibnamefont {Tenkanen}}, \bibinfo {author} {\bibfnamefont {K.}~\bibnamefont {Tuominen}},\ and\ \bibinfo {author} {\bibfnamefont {V.}~\bibnamefont {Vaskonen}},\ }\href {https://doi.org/10.1142/S0217751X1730023X} {\bibfield  {journal} {\bibinfo  {journal} {Int. J. Mod. Phys. A}\ }\textbf {\bibinfo {volume} {32}},\ \bibinfo {pages} {1730023} (\bibinfo {year} {2017})},\ \Eprint {https://arxiv.org/abs/1706.07442} {arXiv:1706.07442 [hep-ph]} \BibitemShut {NoStop}%
\bibitem [{\citenamefont {Hall}\ \emph {et~al.}(2010)\citenamefont {Hall}, \citenamefont {Jedamzik}, \citenamefont {March-Russell},\ and\ \citenamefont {West}}]{Hall:2009bx}%
  \BibitemOpen
  \bibfield  {author} {\bibinfo {author} {\bibfnamefont {L.~J.}\ \bibnamefont {Hall}}, \bibinfo {author} {\bibfnamefont {K.}~\bibnamefont {Jedamzik}}, \bibinfo {author} {\bibfnamefont {J.}~\bibnamefont {March-Russell}},\ and\ \bibinfo {author} {\bibfnamefont {S.~M.}\ \bibnamefont {West}},\ }\href {https://doi.org/10.1007/JHEP03(2010)080} {\bibfield  {journal} {\bibinfo  {journal} {JHEP}\ }\textbf {\bibinfo {volume} {03}},\ \bibinfo {pages} {080}},\ \Eprint {https://arxiv.org/abs/0911.1120} {arXiv:0911.1120 [hep-ph]} \BibitemShut {NoStop}%
\bibitem [{\citenamefont {Racco}\ and\ \citenamefont {Riotto}(2023)}]{Racco:2022svs}%
  \BibitemOpen
  \bibfield  {author} {\bibinfo {author} {\bibfnamefont {D.}~\bibnamefont {Racco}}\ and\ \bibinfo {author} {\bibfnamefont {A.}~\bibnamefont {Riotto}},\ }\href {https://doi.org/10.1088/1475-7516/2023/01/020} {\bibfield  {journal} {\bibinfo  {journal} {JCAP}\ }\textbf {\bibinfo {volume} {01}},\ \bibinfo {pages} {020}},\ \Eprint {https://arxiv.org/abs/2211.08719} {arXiv:2211.08719 [hep-ph]} \BibitemShut {NoStop}%
\bibitem [{\citenamefont {Strumia}(2023)}]{Strumia:2022qvj}%
  \BibitemOpen
  \bibfield  {author} {\bibinfo {author} {\bibfnamefont {A.}~\bibnamefont {Strumia}},\ }\href {https://doi.org/10.1007/JHEP03(2023)042} {\bibfield  {journal} {\bibinfo  {journal} {JHEP}\ }\textbf {\bibinfo {volume} {03}},\ \bibinfo {pages} {042}},\ \Eprint {https://arxiv.org/abs/2211.08359} {arXiv:2211.08359 [hep-ph]} \BibitemShut {NoStop}%
\bibitem [{\citenamefont {Sirko}(2005)}]{Sirko:2005uz}%
  \BibitemOpen
  \bibfield  {author} {\bibinfo {author} {\bibfnamefont {E.}~\bibnamefont {Sirko}},\ }\href {https://doi.org/10.1086/497090} {\bibfield  {journal} {\bibinfo  {journal} {Astrophys. J.}\ }\textbf {\bibinfo {volume} {634}},\ \bibinfo {pages} {728} (\bibinfo {year} {2005})},\ \Eprint {https://arxiv.org/abs/astro-ph/0503106} {arXiv:astro-ph/0503106} \BibitemShut {NoStop}%
\bibitem [{\citenamefont {Gnedin}\ \emph {et~al.}(2011)\citenamefont {Gnedin}, \citenamefont {Kravtsov},\ and\ \citenamefont {Rudd}}]{Gnedin:2011kj}%
  \BibitemOpen
  \bibfield  {author} {\bibinfo {author} {\bibfnamefont {N.~Y.}\ \bibnamefont {Gnedin}}, \bibinfo {author} {\bibfnamefont {A.~V.}\ \bibnamefont {Kravtsov}},\ and\ \bibinfo {author} {\bibfnamefont {D.~H.}\ \bibnamefont {Rudd}},\ }\href {https://doi.org/10.1088/0067-0049/194/2/46} {\bibfield  {journal} {\bibinfo  {journal} {Astrophys. J. Suppl.}\ }\textbf {\bibinfo {volume} {194}},\ \bibinfo {pages} {46} (\bibinfo {year} {2011})},\ \Eprint {https://arxiv.org/abs/1104.1428} {arXiv:1104.1428 [astro-ph.CO]} \BibitemShut {NoStop}%
\bibitem [{\citenamefont {Baldauf}\ \emph {et~al.}(2011)\citenamefont {Baldauf}, \citenamefont {Seljak}, \citenamefont {Senatore},\ and\ \citenamefont {Zaldarriaga}}]{Baldauf:2011bh}%
  \BibitemOpen
  \bibfield  {author} {\bibinfo {author} {\bibfnamefont {T.}~\bibnamefont {Baldauf}}, \bibinfo {author} {\bibfnamefont {U.}~\bibnamefont {Seljak}}, \bibinfo {author} {\bibfnamefont {L.}~\bibnamefont {Senatore}},\ and\ \bibinfo {author} {\bibfnamefont {M.}~\bibnamefont {Zaldarriaga}},\ }\href {https://doi.org/10.1088/1475-7516/2011/10/031} {\bibfield  {journal} {\bibinfo  {journal} {JCAP}\ }\textbf {\bibinfo {volume} {10}},\ \bibinfo {pages} {031}},\ \Eprint {https://arxiv.org/abs/1106.5507} {arXiv:1106.5507 [astro-ph.CO]} \BibitemShut {NoStop}%
\bibitem [{\citenamefont {Li}\ \emph {et~al.}(2014)\citenamefont {Li}, \citenamefont {Hu},\ and\ \citenamefont {Takada}}]{Li:2014sga}%
  \BibitemOpen
  \bibfield  {author} {\bibinfo {author} {\bibfnamefont {Y.}~\bibnamefont {Li}}, \bibinfo {author} {\bibfnamefont {W.}~\bibnamefont {Hu}},\ and\ \bibinfo {author} {\bibfnamefont {M.}~\bibnamefont {Takada}},\ }\href {https://doi.org/10.1103/PhysRevD.89.083519} {\bibfield  {journal} {\bibinfo  {journal} {Phys. Rev. D}\ }\textbf {\bibinfo {volume} {89}},\ \bibinfo {pages} {083519} (\bibinfo {year} {2014})},\ \Eprint {https://arxiv.org/abs/1401.0385} {arXiv:1401.0385 [astro-ph.CO]} \BibitemShut {NoStop}%
\bibitem [{\citenamefont {Hu}\ \emph {et~al.}(2016)\citenamefont {Hu}, \citenamefont {Chiang}, \citenamefont {Li},\ and\ \citenamefont {LoVerde}}]{Hu:2016ssz}%
  \BibitemOpen
  \bibfield  {author} {\bibinfo {author} {\bibfnamefont {W.}~\bibnamefont {Hu}}, \bibinfo {author} {\bibfnamefont {C.-T.}\ \bibnamefont {Chiang}}, \bibinfo {author} {\bibfnamefont {Y.}~\bibnamefont {Li}},\ and\ \bibinfo {author} {\bibfnamefont {M.}~\bibnamefont {LoVerde}},\ }\href {https://doi.org/10.1103/PhysRevD.94.023002} {\bibfield  {journal} {\bibinfo  {journal} {Phys. Rev. D}\ }\textbf {\bibinfo {volume} {94}},\ \bibinfo {pages} {023002} (\bibinfo {year} {2016})},\ \Eprint {https://arxiv.org/abs/1605.01412} {arXiv:1605.01412 [astro-ph.CO]} \BibitemShut {NoStop}%
\bibitem [{\citenamefont {Zegeye}\ \emph {et~al.}(2022)\citenamefont {Zegeye}, \citenamefont {Inomata},\ and\ \citenamefont {Hu}}]{Zegeye:2021yml}%
  \BibitemOpen
  \bibfield  {author} {\bibinfo {author} {\bibfnamefont {D.}~\bibnamefont {Zegeye}}, \bibinfo {author} {\bibfnamefont {K.}~\bibnamefont {Inomata}},\ and\ \bibinfo {author} {\bibfnamefont {W.}~\bibnamefont {Hu}},\ }\href {https://doi.org/10.1103/PhysRevD.105.103535} {\bibfield  {journal} {\bibinfo  {journal} {Phys. Rev. D}\ }\textbf {\bibinfo {volume} {105}},\ \bibinfo {pages} {103535} (\bibinfo {year} {2022})},\ \Eprint {https://arxiv.org/abs/2112.05190} {arXiv:2112.05190 [astro-ph.CO]} \BibitemShut {NoStop}%
\bibitem [{\citenamefont {Inomata}\ \emph {et~al.}(2023)\citenamefont {Inomata}, \citenamefont {Lee},\ and\ \citenamefont {Hu}}]{Inomata:2023faq}%
  \BibitemOpen
  \bibfield  {author} {\bibinfo {author} {\bibfnamefont {K.}~\bibnamefont {Inomata}}, \bibinfo {author} {\bibfnamefont {H.}~\bibnamefont {Lee}},\ and\ \bibinfo {author} {\bibfnamefont {W.}~\bibnamefont {Hu}},\ }\href {https://doi.org/10.1088/1475-7516/2023/08/021} {\bibfield  {journal} {\bibinfo  {journal} {JCAP}\ }\textbf {\bibinfo {volume} {08}},\ \bibinfo {pages} {021}},\ \Eprint {https://arxiv.org/abs/2304.10559} {arXiv:2304.10559 [astro-ph.CO]} \BibitemShut {NoStop}%
\bibitem [{\citenamefont {Binder}\ \emph {et~al.}(2017)\citenamefont {Binder}, \citenamefont {Bringmann}, \citenamefont {Gustafsson},\ and\ \citenamefont {Hryczuk}}]{Binder:2017rgn}%
  \BibitemOpen
  \bibfield  {author} {\bibinfo {author} {\bibfnamefont {T.}~\bibnamefont {Binder}}, \bibinfo {author} {\bibfnamefont {T.}~\bibnamefont {Bringmann}}, \bibinfo {author} {\bibfnamefont {M.}~\bibnamefont {Gustafsson}},\ and\ \bibinfo {author} {\bibfnamefont {A.}~\bibnamefont {Hryczuk}},\ }\href {https://doi.org/10.1103/PhysRevD.96.115010} {\bibfield  {journal} {\bibinfo  {journal} {Phys. Rev. D}\ }\textbf {\bibinfo {volume} {96}},\ \bibinfo {pages} {115010} (\bibinfo {year} {2017})},\ \bibinfo {note} {\href{https://link.aps.org/doi/10.1103/PhysRevD.101.099901}{[Erratum: Phys.Rev.D 101, 099901 (2020)]}},\ \Eprint {https://arxiv.org/abs/1706.07433} {arXiv:1706.07433 [astro-ph.CO]} \BibitemShut {NoStop}%
\bibitem [{\citenamefont {Berlin}\ \emph {et~al.}(2014)\citenamefont {Berlin}, \citenamefont {Hooper},\ and\ \citenamefont {McDermott}}]{Berlin:2014tja}%
  \BibitemOpen
  \bibfield  {author} {\bibinfo {author} {\bibfnamefont {A.}~\bibnamefont {Berlin}}, \bibinfo {author} {\bibfnamefont {D.}~\bibnamefont {Hooper}},\ and\ \bibinfo {author} {\bibfnamefont {S.~D.}\ \bibnamefont {McDermott}},\ }\href {https://doi.org/10.1103/PhysRevD.89.115022} {\bibfield  {journal} {\bibinfo  {journal} {Phys. Rev. D}\ }\textbf {\bibinfo {volume} {89}},\ \bibinfo {pages} {115022} (\bibinfo {year} {2014})},\ \Eprint {https://arxiv.org/abs/1404.0022} {arXiv:1404.0022 [hep-ph]} \BibitemShut {NoStop}%
\bibitem [{\citenamefont {Workman}\ and\ \citenamefont {Others}(2022)}]{Workman:2022ynf}%
  \BibitemOpen
  \bibfield  {author} {\bibinfo {author} {\bibfnamefont {R.~L.}\ \bibnamefont {Workman}}\ and\ \bibinfo {author} {\bibnamefont {Others}} (\bibinfo {collaboration} {Particle Data Group}),\ }\href {https://doi.org/10.1093/ptep/ptac097} {\bibfield  {journal} {\bibinfo  {journal} {PTEP}\ }\textbf {\bibinfo {volume} {2022}},\ \bibinfo {pages} {083C01} (\bibinfo {year} {2022})}\BibitemShut {NoStop}%
\bibitem [{\citenamefont {Gondolo}\ and\ \citenamefont {Gelmini}(1991)}]{GondoloCosmicAbundancesStable1991}%
  \BibitemOpen
  \bibfield  {author} {\bibinfo {author} {\bibfnamefont {P.}~\bibnamefont {Gondolo}}\ and\ \bibinfo {author} {\bibfnamefont {G.}~\bibnamefont {Gelmini}},\ }\href {https://doi.org/10.1016/0550-3213(91)90438-4} {\bibfield  {journal} {\bibinfo  {journal} {Nucl. Phys. B}\ }\textbf {\bibinfo {volume} {360}},\ \bibinfo {pages} {145} (\bibinfo {year} {1991})}\BibitemShut {NoStop}%
\bibitem [{\citenamefont {Laine}\ and\ \citenamefont {Meyer}(2015)}]{Laine:2015kra}%
  \BibitemOpen
  \bibfield  {author} {\bibinfo {author} {\bibfnamefont {M.}~\bibnamefont {Laine}}\ and\ \bibinfo {author} {\bibfnamefont {M.}~\bibnamefont {Meyer}},\ }\href {https://doi.org/10.1088/1475-7516/2015/07/035} {\bibfield  {journal} {\bibinfo  {journal} {JCAP}\ }\textbf {\bibinfo {volume} {07}},\ \bibinfo {pages} {035}},\ \Eprint {https://arxiv.org/abs/1503.04935} {arXiv:1503.04935 [hep-ph]} \BibitemShut {NoStop}%
\bibitem [{\citenamefont {Borsanyi}\ \emph {et~al.}(2016)\citenamefont {Borsanyi} \emph {et~al.}}]{Borsanyi:2016ksw}%
  \BibitemOpen
  \bibfield  {author} {\bibinfo {author} {\bibfnamefont {S.}~\bibnamefont {Borsanyi}} \emph {et~al.},\ }\href {https://doi.org/10.1038/nature20115} {\bibfield  {journal} {\bibinfo  {journal} {Nature}\ }\textbf {\bibinfo {volume} {539}},\ \bibinfo {pages} {69} (\bibinfo {year} {2016})},\ \Eprint {https://arxiv.org/abs/1606.07494} {arXiv:1606.07494 [hep-lat]} \BibitemShut {NoStop}%
\bibitem [{\citenamefont {Husdal}(2016)}]{Husdal:2016haj}%
  \BibitemOpen
  \bibfield  {author} {\bibinfo {author} {\bibfnamefont {L.}~\bibnamefont {Husdal}},\ }\href {https://doi.org/10.3390/galaxies4040078} {\bibfield  {journal} {\bibinfo  {journal} {Galaxies}\ }\textbf {\bibinfo {volume} {4}},\ \bibinfo {pages} {78} (\bibinfo {year} {2016})},\ \Eprint {https://arxiv.org/abs/1609.04979} {arXiv:1609.04979 [astro-ph.CO]} \BibitemShut {NoStop}%
\bibitem [{\citenamefont {Bringmann}(2009)}]{Bringmann:2009vf}%
  \BibitemOpen
  \bibfield  {author} {\bibinfo {author} {\bibfnamefont {T.}~\bibnamefont {Bringmann}},\ }\href {https://doi.org/10.1088/1367-2630/11/10/105027} {\bibfield  {journal} {\bibinfo  {journal} {New J. Phys.}\ }\textbf {\bibinfo {volume} {11}},\ \bibinfo {pages} {105027} (\bibinfo {year} {2009})},\ \Eprint {https://arxiv.org/abs/0903.0189} {arXiv:0903.0189 [astro-ph.CO]} \BibitemShut {NoStop}%
\bibitem [{\citenamefont {Du}\ \emph {et~al.}(2022)\citenamefont {Du}, \citenamefont {Huang}, \citenamefont {Li}, \citenamefont {Li},\ and\ \citenamefont {Yu}}]{DuRevisitingDarkMatter2022}%
  \BibitemOpen
  \bibfield  {author} {\bibinfo {author} {\bibfnamefont {Y.}~\bibnamefont {Du}}, \bibinfo {author} {\bibfnamefont {F.}~\bibnamefont {Huang}}, \bibinfo {author} {\bibfnamefont {H.-L.}\ \bibnamefont {Li}}, \bibinfo {author} {\bibfnamefont {Y.-Z.}\ \bibnamefont {Li}},\ and\ \bibinfo {author} {\bibfnamefont {J.-H.}\ \bibnamefont {Yu}},\ }\href {https://doi.org/10.1088/1475-7516/2022/04/012} {\bibfield  {journal} {\bibinfo  {journal} {J. Cosmol. Astropart. Phys.}\ }\textbf {\bibinfo {volume} {2022}}\bibinfo  {number} { (04)},\ \bibinfo {pages} {012}}\BibitemShut {NoStop}%
\bibitem [{\citenamefont {Kawasaki}\ \emph {et~al.}(2021)\citenamefont {Kawasaki}, \citenamefont {Nakatsuka}, \citenamefont {Nakayama},\ and\ \citenamefont {Sekiguchi}}]{Kawasaki:2021etm}%
  \BibitemOpen
\bibfield  {number} {  }\bibfield  {author} {\bibinfo {author} {\bibfnamefont {M.}~\bibnamefont {Kawasaki}}, \bibinfo {author} {\bibfnamefont {H.}~\bibnamefont {Nakatsuka}}, \bibinfo {author} {\bibfnamefont {K.}~\bibnamefont {Nakayama}},\ and\ \bibinfo {author} {\bibfnamefont {T.}~\bibnamefont {Sekiguchi}},\ }\href {https://doi.org/10.1088/1475-7516/2021/12/015} {\bibfield  {journal} {\bibinfo  {journal} {JCAP}\ }\textbf {\bibinfo {volume} {12}}\bibfield  {number} {\bibinfo  {number} { (12)},\ \bibinfo {pages} {015}},\ }\Eprint {https://arxiv.org/abs/2105.08334} {arXiv:2105.08334 [astro-ph.CO]} \BibitemShut {NoStop}%
\bibitem [{\citenamefont {Stebbins}(2023)}]{Stebbins}%
  \BibitemOpen
  \bibfield  {author} {\bibinfo {author} {\bibfnamefont {A.}~\bibnamefont {Stebbins}},\ }\Eprint {https://arxiv.org/abs/2311.17379} {arXiv:2311.17379 [astro-ph.CO]}  (\bibinfo {year} {2023})\BibitemShut {NoStop}%
\end{thebibliography}%

\end{document}